\documentclass[twocolumn,floatfix,showpacs,prb,10pt]{revtex4-1}
\usepackage{graphicx}
\usepackage{amsmath}
\usepackage{amssymb}
\usepackage{bm}
\usepackage{hyperref}
\usepackage{multirow}
\usepackage{color}
\usepackage{bbold}

\newcommand{\pd}{{\phantom\dag}}

\begin{document}

\title{Scattering matrix invariants of Floquet topological insulators}

\author{I. C. Fulga}
\affiliation{Department of Condensed Matter Physics, Weizmann Institute of Science, Rehovot 76100, Israel}

\author{M. Maksymenko}
\altaffiliation{On leave from the Institute for Condensed Matter Physics, National Academy of Sciences of Ukraine, L'viv-79011, Ukraine}
\affiliation{Department of Condensed Matter Physics, Weizmann Institute of Science, Rehovot 76100, Israel}

\date{\today}
\begin{abstract}
Similar to static systems, periodically driven systems can host a variety of topologically non-trivial phases. Unlike the case of static Hamiltonians, the topological indices of bulk Floquet bands may fail to describe the presence and robustness of edge states, prompting the search for new invariants. We develop a unified description of topological phases and their invariants in driven systems, by using scattering theory. We show that scattering matrix invariants correctly describe the topological phase, even when all bulk Floquet bands are trivial. Additionally, we use scattering theory to introduce and analyze new periodically driven phases, such as \emph{weak topological Floquet insulators}, for which invariants were previously unknown. We highlight some of their similarities with static systems, including robustness to disorder, as well as some of the features unique to driven systems, showing that the weak phase may be destroyed by breaking translational symmetry not in space, but in time.
\end{abstract}
% \pacs{...}
\maketitle

\section{Introduction}
\label{sec:intro}

Topologically non-trivial phases are characterized by the simultaneous presence of an insulating bulk and of robust, conducting edge states.\cite{Hasan2010, Qi2011} The boundaries are protected against localization by the system's bulk properties: its symmetries and the presence of a mobility gap. This link is termed bulk-boundary correspondence and it enables defining topological invariants, $\mathbb{Z}$ or $\mathbb{Z}_2$ valued quantities computed from the bulk system, which determine the presence and number of protected gapless modes at the boundary.

The number and nature of topological invariants has been extensively studied for static Hamiltonians. For non-interacting systems, there exist classifications of symmetry protected topological insulators and superconductors.\cite{Kitaev2009, Schnyder2009, Slager2012, Jadaun2013, Chiu2013, Zhang2013, Benalcazar2014, Morimoto2013, Diez2015} Their boundary states can be protected not only by the three fundamental symmetries of the system: time-reversal, particle-hole, and chiral symmetry, but also by symmetries of the underlying lattice. The latter are known as weak topological insulators in the case of translation symmetry or topological crystalline insulators for point group symmetries (reflection, rotation, etc.), and can exist both in two and three space dimensions.\cite{Moore2007, Roy2009, Fu2007, Fu2011, Hsieh2012, Seroussi2014, Diez2014a} Recent experiments have shown evidence for the existence and robustness of both.\cite{Tanaka2012, Dziawa2012, Xu2012, Pauly2015}

Beyond static Hamiltonians, the topological phases of periodically driven systems have recently become the focus of a wide range of research.\cite{Cayssol2013, Bukov2015, Goldman2014} Part of the appeal of so-called Floquet topological insulators is the possibility to change a system's properties by altering the driving field.\cite{Lindner2011, Wang2013} While the physical properties of time-independent systems are mostly fixed during the fabrication process, a Floquet system can in principle host a variety of topological phases as a function of an external, time-periodic perturbation. Examples include phases hosting Floquet Majorana fermions,\cite{Jiang2011, Kundu2013, Reynoso2013, Thakurathi2013} as well as robust chiral,\cite{Rechtsman2013, Rudner2013, Usaj2014, Jotzu2014, Zheng2014} helical,\cite{Lindner2011, Carpentier2015} or counter-propagating edge modes.\cite{Lababidi2014, Zhou2014, Zhou2014a, Quelle201527, Ho2014} Possible extensions of these ideas to interacting systems are also actively discussed.\cite{Grushin2014, Abanin:2015aa}

In spite of this intense activity, the topological nature of periodically driven systems is much less understood as compared to their time-independent counterparts. Unlike wavefunctions of static systems, Floquet states are obtained from the unitary time-evolution operator over one driving period, the so-called Floquet operator, ${\cal F}$.
Each of its eigenstates accumulates a phase factor during one period $T$ of the time evolution, as ${\cal F}|\psi\rangle = \exp(-i\varepsilon T/\hbar)|\psi\rangle$, with $\varepsilon$ referred to as quasi-energy. While the Floquet states and their associated quasi-energies are in many ways analogous to the eigenstates and energies of static systems, the periodic nature of the driving means that $\varepsilon$ is only defined modulo $2\pi\hbar/T$.
The periodicity of the resulting Brillouin zone (BZ) in quasi-energy can lead to situations in which the original topological invariants fail to correctly capture the system's behavior at an edge. Indeed, in Ref.~\onlinecite{Rudner2013} it was shown that robust chiral edge states can form in a system where all bulk bands are trivial, since the Floquet operator in the bulk equals the unit matrix, ${\cal F} = \mathbb{1}$.

Several works have made progress towards a full classification of Floquet topological insulators.\cite{Kitagawa2010, Nathan2015} In some cases, novel invariants have been formulated, which take into account the system's driven nature. For strong two-dimensional (2d) Floquet topological insulators, invariants have been found, both in the presence and absence of time-reversal symmetry.\cite{Rudner2013, Carpentier2015} Weak topological effects, however, have remained largely unexplored. Among the few examples to date, 2d chiral-symmetric systems hosting anomalous counter-propagating edge modes have been reported in Refs.~\onlinecite{Lababidi2014, Zhou2014, Zhou2014a, Ho2014}. In these works the existence and robustness to disorder of edge modes was identified, but the topological index responsible for their presence has remained up to now unknown.

To accurately predict the possible non-trivial phases, the properties of Floquet topological insulators -- both weak and strong -- need to be treated within a unified framework. To this end, rather than focusing on their difference with respect to static systems, we take advantage of their similarities instead. In both types of phases, topological properties manifest themselves at the (quasi-)Fermi level, in the form of robust conducting edge modes, which exhibit similar forms of spectral flow. As such, their non-trivial phases can be characterized in a unified manner, by using scattering theory.

Previous works have considered the quantum transport properties of Floquet systems.\cite{Moskalets2002, Kitagawa2011, FoaTorres2014, Perez-Piskunow2014} The scattering problem is typically defined in terms of a driven sample connected to multiple metallic electrodes and the scattering matrix relates the amplitudes of asymptotic scattering states in the metallic leads. In contrast, our main focus is not studying transport properties, but establishing a topological classification of Floquet systems in terms of the scattering matrix. To this end, rather than attaching metallic electrodes, we impose absorbing boundary conditions on a given sample to define a simplified, yet fictitious scattering problem.\cite{Tajic2005} As we will show, the resulting scattering matrix enables to fully determine the topological properties of periodically driven systems.

The Fermi level scattering matrix has been used to formulate both strong and weak topological invariants for time-independent Hamiltonians\cite{Fulga2011, Fulga2012} as well as 1d quantum walks,\cite{Tarasinski2014} having the advantage of being naturally tailored to the study of disordered systems. In this work, we extend this approach to periodically driven systems, showing different examples of its application. 
Specifically, we formulate novel topological invariants for the models of Refs.~\onlinecite{Lababidi2014, Ho2014}, based on the constraints imposed on the scattering matrix by the symmetries of the system. Furthermore, we show that the original scattering matrix invariants developed to characterize strong and weak \emph{static} topological insulators can be readily applied to driven systems, even in cases in which all bulk Floquet bands are trivial. This unified approach mitigates the need for new topological indices, enabling the study of a wide class of both static and driven phases using the same invariant expressions.
As an example, we turn to the model studied in Ref.~\onlinecite{Kitagawa2010} and reveal a richer topological structure consisting of both strong and weak phases. We showcase some features of weak phases which are unique to periodically driven systems, such as the possibility of gapping out the edge modes by breaking translation symmetry in time.

The rest of our work is organized as follows. In Section \ref{sec:chiral} we study the chiral-symmetric Floquet system of Refs.~\onlinecite{Lababidi2014, Ho2014}. We determine the scattering matrix invariant and, based on its structure, predict the localization behavior of edge states in the presence of disorder. We find non-trivial phases of a $\mathbb{Z}_2$ (as opposed to $\mathbb{Z}$) nature, and conclude that the edge states are not robust to disorder which does not preserve chiral symmetry, contrary to previous reports. We confirm this expectation by performing numerical simulations. In Section \ref{sec:phs} we turn to the particle-hole symmetric Floquet topological insulator of Ref.~\onlinecite{Kitagawa2010}, which hosts both strong and weak topological phases. We find that the original, strong and weak scattering matrix invariants accurately describe the edge state robustness, even though all bulk bands are trivial. We conclude in Section \ref{sec:conc}.

\section{Chiral Symmetric Driven System}
\label{sec:chiral}

To date, there are few studies of weak topological effects in periodically driven systems. Refs.~\onlinecite{Lababidi2014, Ho2014} have observed this behavior in 2d systems possessing chiral symmetry. In this Section, we turn to one such model, the \emph{kicked quantum Hall} system, and use scattering theory to determine its topological invariant.

The model describes spinless fermions on a square lattice (lattice constant $a=1$), where the position of each site is given by the vector ${\bf r} = n_x \hat{x} + n_y \hat{y}$, with $n_{x, y}$ integers and $\hat{x}$, $\hat{y}$ unit vectors pointing in the $x$- and $y$-directions, respectively.

The tight binding Hamiltonian

\begin{equation}\label{eq:Hkqh}
\begin{split}
 H_{\rm KQH} = & \frac{J_x}{2} \sum_{n_x, n_y} |n_x+1, n_y\rangle\langle n_x, n_y| + \\
 & \frac{J_y}{2} e^{in_x\alpha}|n_x, n_y+1\rangle\langle n_x, n_y| + {\rm h.c.},
\end{split}
\end{equation}
is expressed in terms of states $|n_x, n_y\rangle$ on lattice sites indexed by $(n_x, n_y)$. Here, $J_{x,y}$ are the nearest neighbor hopping amplitudes in the $x$- and $y$-directions. The system is placed in a uniform magnetic field, with $\alpha$ modeling the flux threaded in each plaquette. 

In the static case, Eq.~\eqref{eq:Hkqh} is the well-studied Hofstadter model,\cite{Hofstadter1976} showing a fractal pattern of gapped, Chern-insulating phases. As in Refs.~\onlinecite{Lababidi2014, Ho2014}, we take $J_{x,y}$ to be periodic functions of time, with $J_x(t) = J_x$ and $J_y(t) = J_y \sum_m \delta(t-m)$, $m\in\mathbb{Z}$. In other words, the hopping in the $x$-direction is kept constant, while the coupling in the $y$ direction is turned on periodically with a period $T=1$, but only at discrete times $t=m$.

When the system is infinite in one or more directions, or alternatively in the presence of periodic boundary conditions (PBC), momentum becomes a good quantum number, enabling us to express Eq.~\eqref{eq:Hkqh} in reciprocal space. The time-dependent momentum-space Hamiltonian becomes
\begin{equation}\label{eq:Hkqhk}
 {\cal H}_{\rm KQH} = J_x \cos(k_x) + J_y \cos(k_y - n_x \alpha)\sum_m \delta(t-m),
\end{equation}
where $n_x$ labels sites in the magnetic unit cell.

As long as there are no on-site potentials, the model shows chiral symmetry, expressed as a staggered gauge transformation. The chiral symmetry operator changes the sign of the wavefunction on one of the two sublattices
\begin{equation}\label{eq:chiralsym}
 \Gamma : |n_x,n_y\rangle \to (-1)^{n_x+n_y}|n_x, n_y\rangle.
\end{equation}
It is a unitary operator, $\Gamma^\dag\Gamma=\Gamma^2=1$. In real space it anti-commutes with the Hamiltonian, $\Gamma H_{\rm KQH} \Gamma = - H_{\rm KQH}$, while in reciprocal space it amounts to a $\pi$ shift in momentum,

\begin{equation}
\begin{split}
\Gamma {\cal H}_{\rm KQH}(k_x, k_y) \Gamma & = {\cal H}_{\rm KQH}(k_x+\pi, k_y+\pi) \\ &=-{\cal H}_{\rm KQH}(k_x, k_y).
\end{split}
\end{equation}

We set $\hbar=1$ and define the Floquet operator as propagating states from $t=0.5$ to $t=1.5$:
\begin{equation}\label{eq:Fdef}
 {\cal F}_{\rm KQH} = \exp \left( -i\int_{0.5}^{1.5} {\cal H}_{\rm KQH}(t)\,dt \right)
\end{equation}
In this so-called symmetric time frame,\cite{Asboth2013, Asboth2014} Eq.~\eqref{eq:Fdef} can be written as
\begin{equation}\label{eq:Fkqh}
 {\cal F}_{\rm KQH} = e^{-i\frac{J_x}{2}\cos(k_x)} e^{-iJ_y\cos(k_y - n_x \alpha)} e^{-i\frac{J_x}{2}\cos(k_x)}.
\end{equation}
Owing to the chiral symmetry of the periodically modulated Hamiltonian \eqref{eq:Hkqhk}, the Floquet operator \eqref{eq:Fkqh} also shows chiral symmetry, expressed as
\begin{equation}\label{eq:GammaF}
 \Gamma_{\cal F} =\exp(in_x\pi)\,\exp(in_y\pi).
\end{equation}
While on the Hamiltonian level chiral symmetry amounts to changing the sign of the eigenvalues, in Floquet language it reverses the direction of the time-evolution
\begin{equation}\label{eq:Fchiralsym}
 {\Gamma^\pd_{\cal F}}  {\cal F}_{\rm KQH} \Gamma^\pd_{\cal F} = {\cal F}_{\rm KQH}^\dag.
\end{equation}

\begin{figure*}[htb]
 \includegraphics[width=\textwidth]{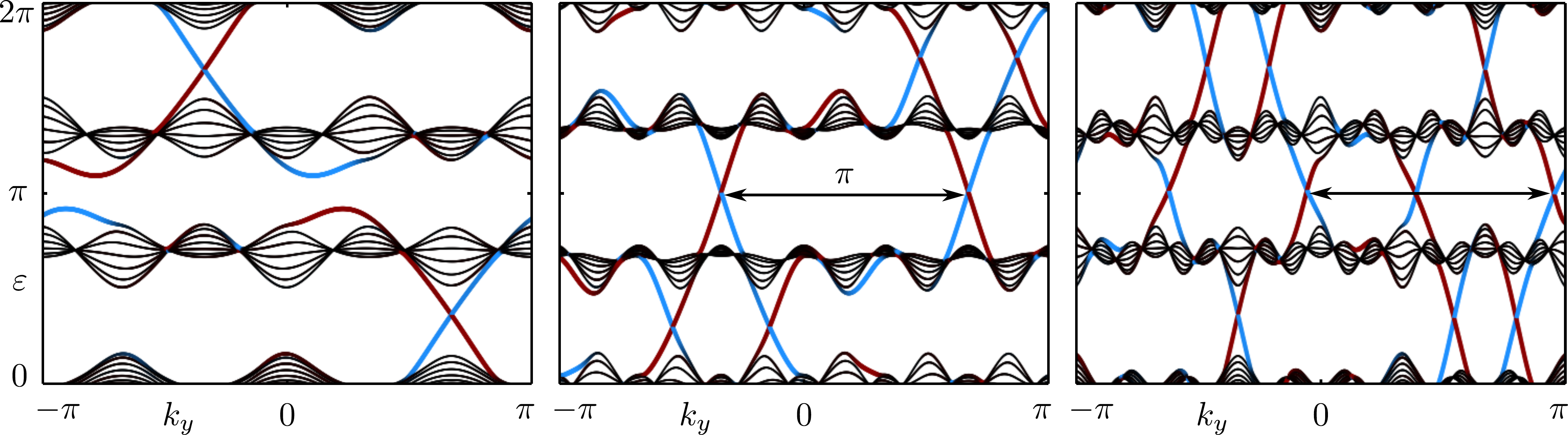}
 \caption{Bandstructure of the Floquet operator \eqref{eq:Fkqh} in an infinite strip geometry (infinite in $y$, $L=21$ sites in the $x$-direction). Parameters are $\alpha=2\pi/3$, $J_x=2\pi/3$, and $J_y=\pi,2\pi,3\pi$, in the left, middle, and right panels, respectively. Red (dark gray) indicates states localized on the left boundary and blue (light gray) states on the right. From left to right, the bulk gap at $\varepsilon=\pi$ shows zero, one, or two pairs of counter propagating edge modes, separated by a momentum difference $\Delta k_y=\pi$ (horizontal arrows).\label{fig:chiralbands}}
\end{figure*}

We set $\alpha=2\pi/3$, such that the system has three bulk Floquet bands. In the following we consider Eq.~\eqref{eq:Fkqh} either in a strip geometry (infinite in $y$, finite in $x$), or discretized on a square lattice of $L\times W$ sites. All numerical results are obtained for tight-binding models defined using the Kwant code.\cite{Groth2014, sourcecode}

Typical banstructures obtained by diagonalizing the Floquet operator ${\cal F}_{\rm KQH}|\psi\rangle = \exp(-i\varepsilon)|\psi\rangle$ in an infinite strip geometry are shown in Fig.~\ref{fig:chiralbands}.
As a function of the parameters $J_{x,y}$, the model exhibits a variety of topological phases. The top and bottom quasi-energy gaps of Fig.~\ref{fig:chiralbands} ($\varepsilon=\pi/2$ and $\varepsilon=3 \pi/2$) show one or more edge modes with a non-zero net chirality, a feature reminiscent of the quantum Hall effect. The boundary modes occur on all edges, irrespective of orientation, and have been understood\cite{Ho2014} in terms of the Floquet winding number of Ref.~\onlinecite{Rudner2013}.

We focus on edge modes in the gap at $\varepsilon=\pi$ which have no net chirality, but come in counter-propagating pairs. Each pair is separated by a momentum difference of $\pi$. Furthermore, unlike boundary states in the top and bottom gaps, the counter-propagating modes depend on the orientation of the edge, similar to weak topological insulators.
For the values of $J_{x,y}$ used in the left panel of Fig.~\ref{fig:chiralbands} there are no gapless edge states on boundaries parallel to either the $x$- or $y$-direction. In the middle panel both types of boundary show one pair of edge modes, while in the right panel 
there are two pairs on the edge parallel to $y$, and none along $x$. There is to date no known topological invariant responsible for the presence and robustness of these edge states.

The presence of counter-propagating modes has been previously identified as being a consequence of chiral symmetry. Indeed, in a strip geometry the relation \eqref{eq:Fchiralsym} becomes
\begin{equation}\label{eq:Fchiralstrip}
 {\Gamma^\pd_{\cal F}}  {\cal F}_{\rm KQH}(k_y) \Gamma^\pd_{\cal F} = {\cal F}_{\rm KQH}^\dag(k_y-\pi),
\end{equation}
such that for each state at energy $\varepsilon$ and momentum $k_y$ there must be a state at $-\varepsilon$ and $k_y-\pi$. Therefore, edge modes at $\varepsilon=\pm\pi$ necessarily come in pairs. In the following we will determine the nature of this topological phase as well as the explicit form of its topological invariant.

\begin{figure}[tb]
 \includegraphics[width=\columnwidth]{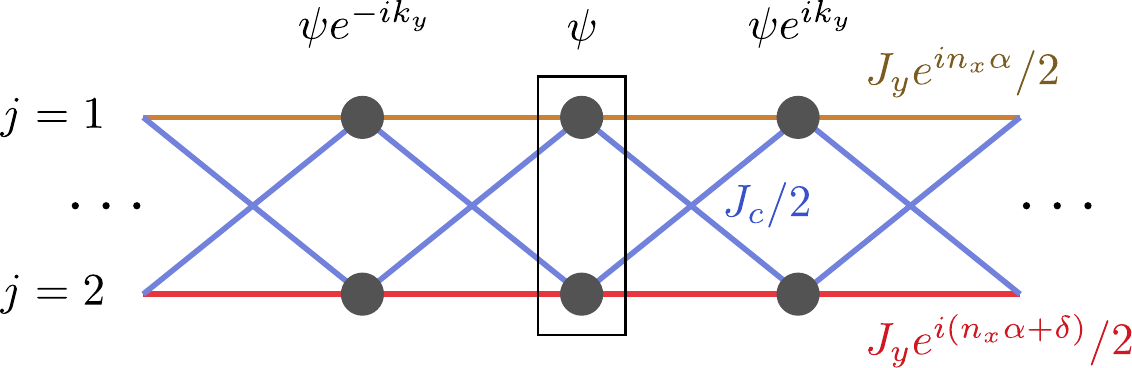}
 \caption{Side view of the Hamiltonian \eqref{eq:2xHkqh}, consisting of two copies of \eqref{eq:Hkqh} labeled by $j=1,2$ (top and bottom layers). One of the three unit cells shown is marked by a box. The coupling within the top and bottom layers are marked with orange and red lines, respectively, while the one between the layers is shown in blue. Notice that chiral symmetry is preserved, as there is no coupling between layers in the same unit cell.\label{fig:kqhx2}}
\end{figure}

\begin{figure*}[htb]
 \includegraphics[width=\textwidth]{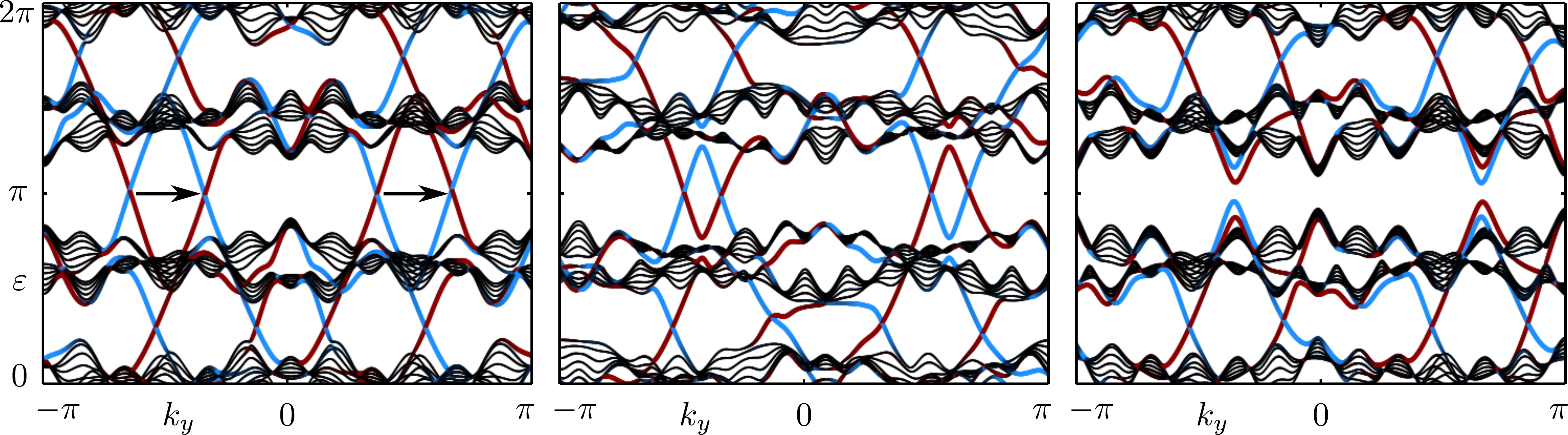}
 \caption{Bandstructures of the model \eqref{eq:2xHkqh} in the same geometry as Fig.~\ref{fig:chiralbands}. We set $\alpha=J_x=2\pi/3$, $J_y=2\pi$, and $J_c=0.6$. In the left, middle, and right panels, $\delta=0.7\pi$, $0.85\pi$, and $0.98\pi$, respectively. Sliding one pair of edge modes on top of the other (horizontal arrows) causes them to annihilate and opens gaps in the spectrum. This indicates they are protected by a $\mathbb{Z}_2$ invariant.\label{fig:slidingedge}}
\end{figure*}

In the absence of disorder momentum is a good quantum number and a single pair of edge states is protected, since the states must be separated by a momentum difference $\Delta k_y=\pi$. Figure \ref{fig:chiralbands} also shows a situation in which there are two pairs of counter-propagating modes at $\varepsilon=\pi$, raising the question of whether this is a distinct phase or whether different pairs of edge modes can annihilate and gap out if they overlap. The answer will determine whether the phase is of a $\mathbb{Z}$ type, with any number of protected edge modes, or of a $\mathbb{Z}_2$ type, where only an odd number of pairs is protected. We test this hypothesis in a controllable fashion, by considering two copies of Eq.~\eqref{eq:Hkqh}, coupled in such a way as to preserve the chiral symmetry \eqref{eq:chiralsym} of the combined system (see Fig.~\ref{fig:kqhx2}). Each of the two subsystems has one pair of edge modes, but one of them is shifted in momentum relative to the other, $k_y \to k_y + \delta$. This enables us to effectively \emph{slide} one pair of edges in the quasi-BZ by changing the value of $\delta$. In a strip geometry, the Hamiltonian of the combined system has the form
\begin{equation}\label{eq:2xHkqh}
\begin{split}
 H_{\rm 2\times KQH} = \sum_{n_x} \Bigg( & \frac{J_x}{2}  \sum_{j=1,2} |n_{x,j}+1\rangle\langle n_{x,j}| + \\
 & \frac{J_y}{2} e^{in_x\alpha+ik_y}|n_{x,1}\rangle\langle n_{x,1}| + \\
 & \frac{J_y}{2}  e^{in_x\alpha+i(k_y+\delta)}|n_{x,2}\rangle\langle n_{x,2}| + \\
 & \frac{J_c}{2} \sum_{j\neq j'} e^{ik_y} |n_{x,j}\rangle\langle n_{x,j'}| + {\rm h.c.} \Bigg),
\end{split}
\end{equation}
where the index $j^{(\prime)}=1,2$ labels the two copies, and $J_c$ is the time-independent coupling between them. In Fig.~\ref{fig:slidingedge} we show that as the two pairs of edge modes are made to overlap, gaps open in the spectrum. Therefore, there are only two topologically distinct edge mode configurations, one pair of counter-propagating modes versus no edge modes, indicating that the non-trivial phase is protected by a $\mathbb{Z}_2$, as opposed to a $\mathbb{Z}$ invariant.

To construct this invariant as well as test the conducting properties of the edge, we turn to the scattering matrix formalism. 
For a static Hamiltonian discretized on a finite lattice, transport properties are usually obtained by connecting one or more infinite, translationally invariant leads to the system. This enables to compute the scattering matrix, which in the two-lead case reads
\begin{equation}\label{eq:smatrix}
 S = \begin{pmatrix}
      r & t \\
      t' & r'
     \end{pmatrix},
\end{equation}
in terms of the reflection and transmission amplitudes of lead modes, $r^{(\prime)}$ and $t^{(\prime)}$.

For periodically driven systems, the scattering matrix associated to a Floquet operator can be computed in a slightly different fashion.\cite{Fyodorov2000, Ossipov2002, Dahlhaus2011} On a square lattice of $L\times W$ sites, the Floquet operator \eqref{eq:Fkqh}  is a $LW\times LW$ matrix. Rather than attaching infinite leads, we introduce absorbing terminals at the two sides of the system, $n_x=1$ and $n_x=L$. Defining a $2W\times LW$ projector onto these terminals,
\begin{equation}\label{eq:projector}
 P = \left\{
 \begin{aligned}
  & 1 \,\,\,{\rm if}\,\,\, n_x\in \{ 1, L \} \\
  & 0 \,\,\,{\rm otherwise}
 \end{aligned}
 \right.
\end{equation}
we can express the quasi-energy dependent Floquet scattering matrix through the formula
\begin{equation}\label{eq:FloquetS}
 S(\varepsilon) = P \left[ 1-e^{i\varepsilon}{\cal F}(1-P^TP) \right]^{-1}e^{i\varepsilon}{\cal F}P^T,
\end{equation}
where the superscript $T$ stands for transposition. The expression \eqref{eq:FloquetS} can be understood by expanding the inverse matrix in a geometric series, in which each subsequent term describes time-evolution over an additional period. Time evolution stops after the state reaches the absorbing terminals, $P$, and continues otherwise, $1-P^TP$.\cite{Tajic2005} The absorbing terminals only act stroboscopically, at the beginning and end of each time period, such that Eq.~\eqref{eq:FloquetS} describes a simplified, fictitious scattering problem. Nevertheless, the unitarity of the Floquet operator ${\cal F}$ implies that the scattering matrix is also unitary, $S(\varepsilon) S^\dag(\varepsilon)=1$, which can be verified though a direct calculation.

In this setup, the Floquet scattering matrix takes the form of Eq.~\eqref{eq:smatrix}, enabling to compute the transmission from one side of the system to the other, $G = {\rm Tr}\,t^\dag t$.
At $\varepsilon=\pi$, the presence of one pair of counter-propagating edge states leads to a quantized transmission $G=2$, since there is one right- and one left-mover on each edge.

First we identify the constraints imposed by the chiral symmetry on $S$. To this end, we consider a system with an odd number of sites in the $y$ direction ($W$ odd) and apply twisted boundary conditions, as $|n_x, W\rangle = e^{i \phi} |n_x, 0\rangle$. The Floquet operator, and therefore also the scattering matrix, now become functions of the twist angle $\phi$, which plays the same role as momentum in the chiral symmetry constraint of Eq.~\eqref{eq:Fchiralstrip}. Plugging this constraint into the Floquet scattering matrix definition \eqref{eq:FloquetS} leads to a relation

\begin{equation}\label{eq:Schiralsym}
 \Gamma^\pd_S S(\varepsilon, \phi) \Gamma^\pd_S = S^\dag(-\varepsilon, \phi-\pi),
\end{equation}
where $\Gamma_S$ is a $2W\times 2W$ matrix defined by $\Gamma_S = P \Gamma_{\cal F} P^T$. In other words, $\Gamma_S$ changes the sign of one sublattice in the absorbing terminals at $n_x=1,L$.

The relation \eqref{eq:Schiralsym} is reminiscent of that found for chiral symmetric static systems,\cite{Fulga2012} where at zero energy there exists a basis in which the scattering matrix is hermitian. For the periodically driven system studied here, the $\pi$ momentum shift induced on the Floquet level carries over in the scattering matrix description. Therefore, we introduce the chiral basis $\widetilde{S} = \Gamma_SS$, in which the reflection sub-block obeys $\widetilde{r}(\phi) = \widetilde{r}^\dag(\phi-\pi)$ at $\varepsilon=\pi$.
This enables us to formulate a topological invariant, by noting that for this quasi-energy
\begin{equation}\label{eq:detr_chiral}
 \det\,\widetilde{r}(\phi) = {\det}^* \,\widetilde{r}(\phi-\pi).
\end{equation}

\begin{figure}[tb]
 \includegraphics[width=0.6\columnwidth]{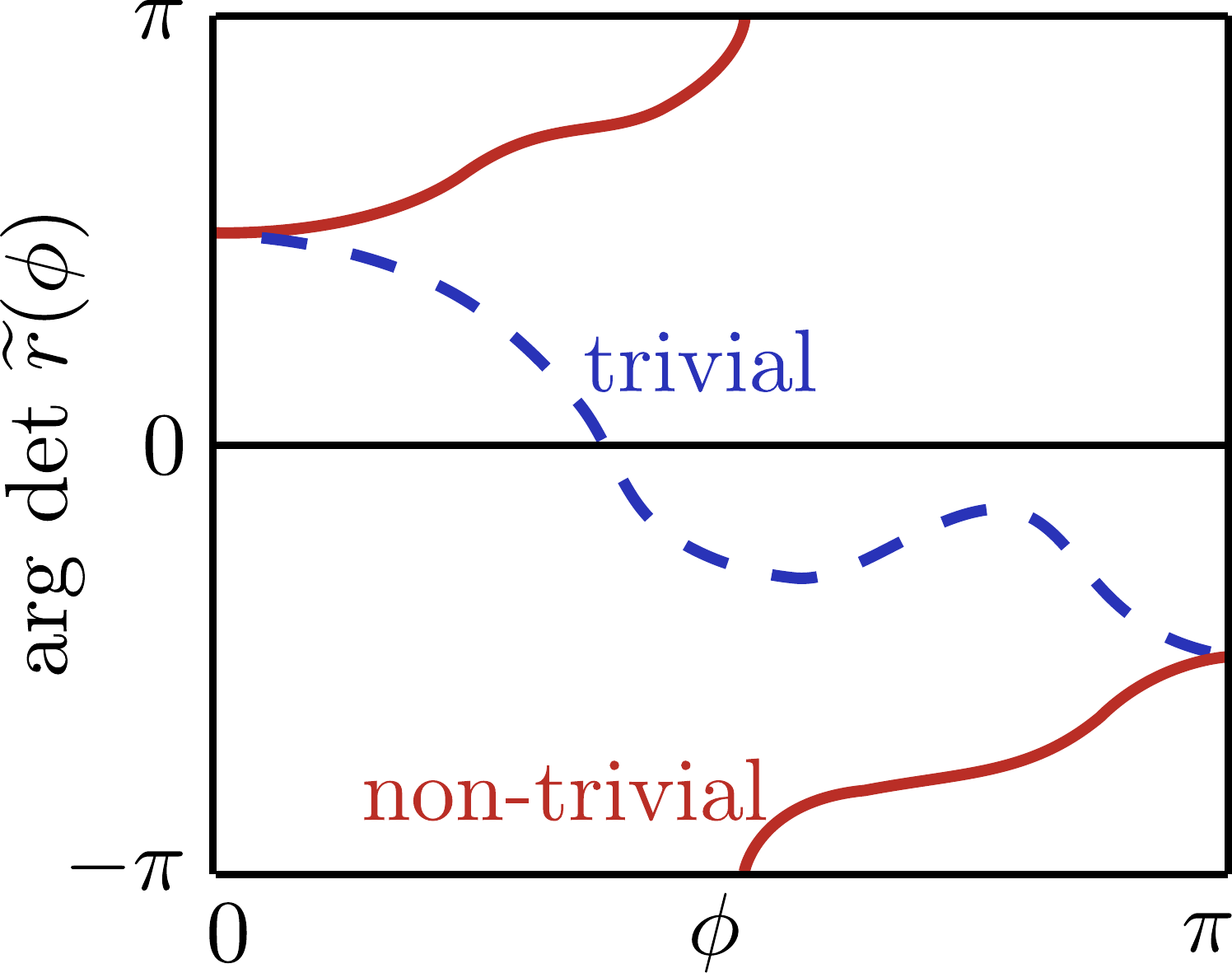}
 \caption{Two possible ways in which the phase of $\det\, \widetilde{r}$ can evolve as the twist angle is advanced from $\phi=0$ to $\phi=\pi$. The solid and dashed lines indicate topologically different scenarios, since they cannot be deformed into each other as long as the bulk mobility gap remains open ($\det \,\widetilde{r}\neq 0$). The presence of protected $\pi$-crossings signals a topologically non-trivial phase in our choice of gauge.\label{fig:Sinvariant}}
\end{figure}

By Eq.~\eqref{eq:detr_chiral}, the complex phase of the determinant of $\widetilde{r}$ must change sign as the twist angle $\phi$ is advanced by $\pi$. Since the complex phase must be a continuous function of $\phi$, only two scenarios are possible. In the interval $\phi\in[0,\pi]$ (half of the effective BZ), either the phase of $\det\,\widetilde{r}$ crosses 0 an odd number of times, or it crosses $\pi$ an odd number of times, as shown in Fig.~\ref{fig:Sinvariant}.

The odd number of 0-crossings or $\pi$-crossings in half of the BZ is a $\mathbb{Z}_2$ topological invariant, and cannot change as long as there is a mobility gap in the bulk: $\det\,\widetilde{r}\neq0$.
While the two scenarios are topologically different, which of them occurs is a matter of definition. In order to preserve the form of the chiral symmetry requirement \eqref{eq:Fchiralstrip}, we have used an odd number of sites in the $y$-direction, making $\widetilde{r}$ a matrix with odd dimensions. Therefore, depending on which of the two sub-lattices changes sign under the application of $\Gamma_S$, the determinant of $\widetilde{r}$ will also change sign, turning 0-crossings into $\pi$-crossings, and vice versa. For definiteness, we chose a gauge in which the chiral basis has $\det \,\Gamma_S=-1$ and identify $\pi$-crossings with a topologically non-trivial phase. Choosing the other sign would imply interchanging the labels of Fig.~\ref{fig:Sinvariant}.

By numerically computing the reflection matrix as a function of twist angle, we have checked that this $\mathbb{Z}_2$ invariant correctly describes the topological phases of the model. Furthermore, the relation \eqref{eq:detr_chiral} still holds when disorder is added to the system, provided the latter does not break chiral symmetry. We therefore expect the edge state transmission to be robust upon the inclusion of random hopping strengths $J_{x,y}$. On the other hand, the system should fully localize if we add random on-site potentials, which violate the chiral symmetry requirement of Eq.~\eqref{eq:Fchiralsym} and \eqref{eq:detr_chiral}.

\begin{figure}[tb]
 \includegraphics[width=0.85\columnwidth]{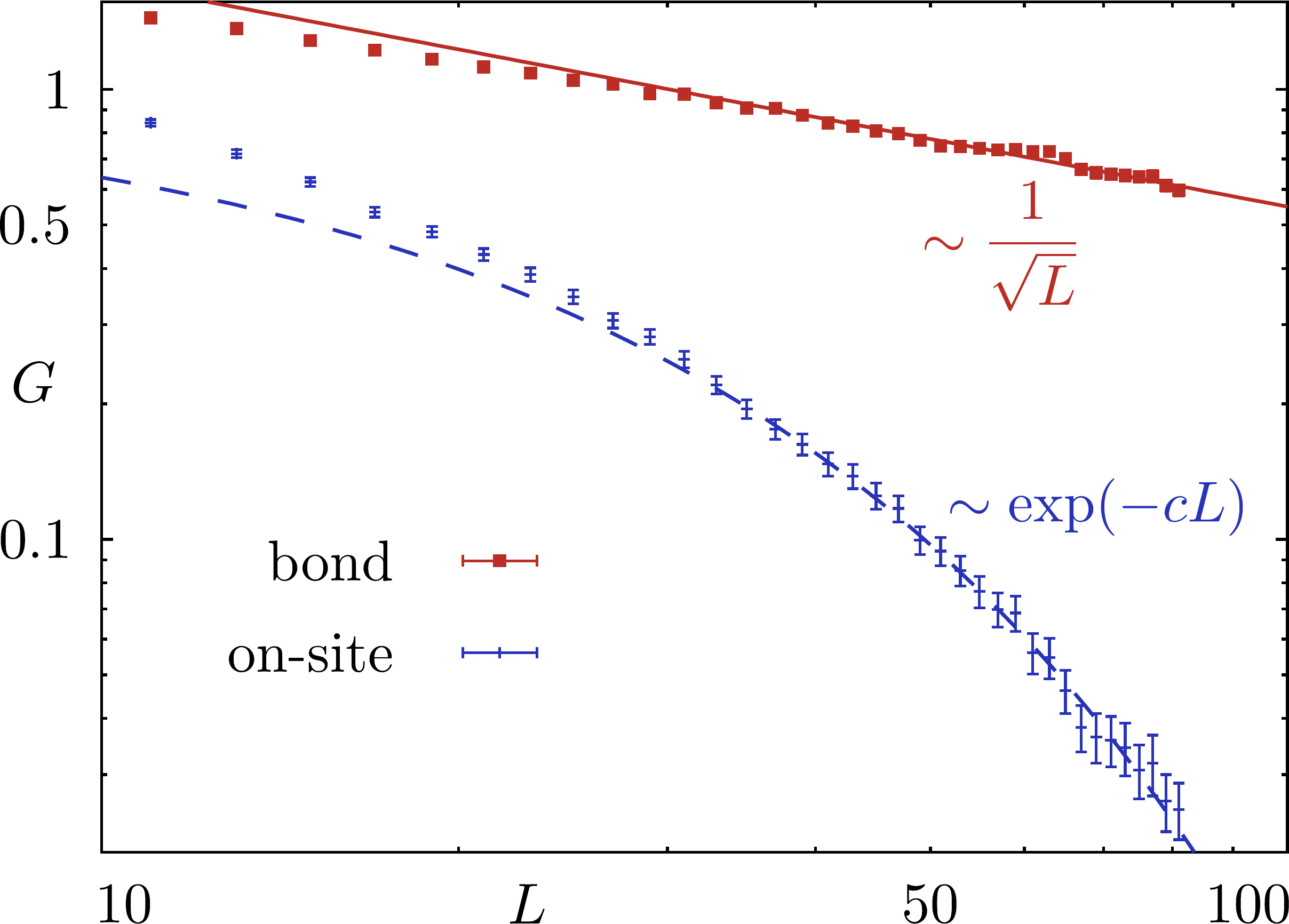}
 \caption{Log-log plot of the edge transmission of model \eqref{eq:Fkqh} at $\varepsilon=\pi$ as a function of system length $L$ for bond (red) and on-site (blue) disorder, with $W=40$, $\alpha=J_x=2\pi/3$, and $J_y=2\pi$. For on-site disorder we have used a disorder strength $U=1.8$, while for bond disorder it was set to $U=0.45$. In each case $600-800$ independent disorder realizations were averaged over. The solid and dashed lines show the expected algebraic and exponential decay of transmission, respectively.\label{fig:chiralscaling}}
\end{figure}

We confirm this numerically by including disorder in one of two different ways. For random bond disorder, we substitute $J_{x,y} \to J_{x,y} (1 + \delta_J)$, with $\delta_J$ drawn independently for each bond from the uniform distribution $[-U, U]$, with $U$ the strength of disorder. In the case of on-site disorder, we introduce a random, time-independent chemical potential term to the Hamiltonian \eqref{eq:Hkqh}, $\delta_{\mu} |n_x,n_y\rangle\langle n_x,n_y|$, where $\delta_\mu$ is drawn independently for each lattice site from the same, uniform distribution. The transmission scaling results of Fig.~\ref{fig:chiralscaling} confirm our expectations. Bond disorder leads to an algebraic decay of the edge transmission, $G\sim 1/\sqrt{L}$, characteristic of two-dimensional statistical topological insulators.\cite{Fulga2014} The edge remains delocalized, being pinned to a one-dimensional critical point,\cite{Brouwer2000, Brouwer2003, Motrunich2001, Gruzberg2005, Diez2014a} and the phase of $\det\, \widetilde{r}$ still shows protected $\pi$-crossings. In contrast, we find that on-site disorder leads to an exponential suppression of edge transmission, $G\sim \exp(-cL)$, $c={\rm const.}$, signaling localization.

Before concluding this Section, we analyze the top and bottom energy gaps appearing in Fig.~\ref{fig:chiralbands}. Away from $\varepsilon=0,\pi$, the system does not obey chiral symmetry, and there is an imbalance in the number of left and right movers on an edge. Since these edge states manifest in the absence of any symmetries, the scattering matrix invariant originally developed to describe the Chern number of static systems\cite{Braeunlich2009, Fulga2012} correctly captures the net chirality of the edge states. By applying twisted boundary conditions as before, we write the topological invariant as the winding number of $\det \,r$
\begin{equation}\label{eq:rchern}
 C_{\rm SM} = \frac{1}{2\pi i}\int_0^{2\pi} d\phi\,\frac{d}{d\phi}\log\,\det\,r(\phi).
\end{equation}
For the parameters in the left and right panels of Fig.~\ref{fig:chiralbands} we find that the top gap has a scattering matrix Chern number $C_{\rm SM}=+1$, while the bottom one has $C_{\rm SM}=-1$. By repeating the calculation for the parameters of the middle panel, we find $C_{\rm SM}=-2$ and $C_{\rm SM}=+2$ for the top and bottom gaps, respectively.

As we have shown, scattering theory enables to fully characterize the topological phases of the model \eqref{eq:Fkqh}. The counter-propagating modes appearing in the gap at $\varepsilon=\pi$ can be understood in terms of an invariant readily obtained by means of a symmetry analysis. Chiral modes present in other gaps are accounted for by the invariant \eqref{eq:rchern}, originally developed in the context of static Chern insulators. In the following Section we will further explore the connection between the scattering matrix invariants of static systems and the non-trivial phases of Floquet topological insulators.

\section{Particle-Hole Symmetric Driven System}
\label{sec:phs}

\begin{figure}
\centering
\includegraphics[width=0.9\columnwidth]{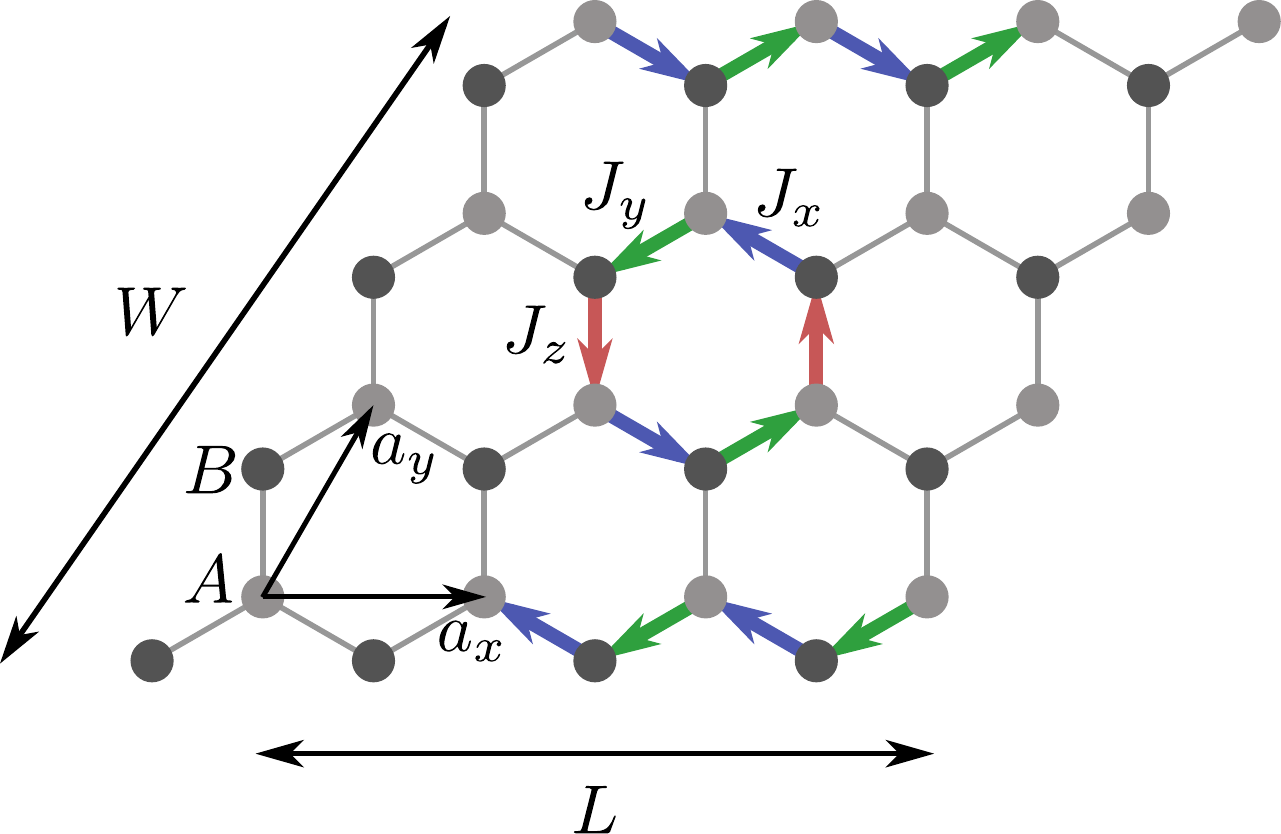}
\caption
{Setup and driving protocol of Hamiltonian \eqref{eq:ph_ham}. We consider an $L\times W$ hexagonal lattice (sublattices $A$ and $B$), with Bravais vectors $a_x$ and $a_y$. When $J_u=0$ only one of the hopping amplitudes $J_{x,y,z}$ (blue, green, and red, respectively) is turned on at any given time. For $J_s=\pi/2$ particles at any site in the bulk return to their original position after two driving periods, while those on the edges form chiral propagating modes, as shown by the arrows.
}
\label{fig:driving_protocol}
\end{figure}

One of the advantages of scattering theory in the study of periodically driven topological phases is that it allows for the study of both time-independent and Floquet topological insulators within the same, unified framework.
In this Section we use the scattering matrix formalism to reveal and characterize the rich structure of a model consisting of both strong and weak topological phases. As we will show, scattering matrix invariants of Floquet systems can take the same form as in their static counterparts, even when bulk Floquet bands fail to capture the non-trivial character of a phase.

We consider an example of particle-hole symmetric driven system, introduced in Ref.~\onlinecite{Kitagawa2010}. It describes spinless fermions on a hexagonal lattice in the presence of nearest neighbor hopping. The unit cell consists of two sites belonging to the two sub-lattices, $A$ and $B$, and the Bravais vectors of the lattice $a_x$ and $a_y$ are defined as in Fig.~\ref{fig:driving_protocol}. 

Setting the lattice constant to 1, the momentum space tight-binding Hamiltonian has the form
\begin{equation}\label{eq:ph_ham_big}
H =\sum_{{\bf{k}}}
\begin{pmatrix}
 c^{\dagger}_{A,{\bf{k}}} c^{\dagger}_{B,{\bf{k}}} 
\end{pmatrix}
{\cal H}
\begin{pmatrix}
c_{A,{\bf{k}}} \\ 
c_{B,{\bf{k}}} 
\end{pmatrix},
\end{equation}
with
\begin{equation}\label{eq:ph_ham}
\begin{split}
{\cal H} =& (J_x \cos(k_y-k_x) + J_y \cos(k_y) + J_z) \sigma_x - \\
& (J_x \sin(k_y-k_x) + J_y \sin(k_y)) \sigma_y.
\end{split}
\end{equation}

Here, $c^\dag_{A/B,{\bf{k}}}$, $c^\pd_{A/B,{\bf{k}}}$ are fermionic creation and annihilation operators on the $A$ and $B$ sub-lattices and $J_{x,y,z}$ are independent hopping amplitudes describing three different types of bonds, as shown in Fig.~\ref{fig:driving_protocol}. The Pauli matrices $\sigma_i$ appearing in ${\cal H}$ parametrize the sub-lattice degree of freedom.

In the absence of driving, the Hamiltonian \eqref{eq:ph_ham} belongs to class BDI in the Altland-Zirnbauer classification.\cite{Altland1997} It is characterized by time-reversal, particle-hole, as well as chiral symmetries, all of which square to $+1$. The corresponding operators are ${\cal T}={\cal K}$, ${\cal P}=\sigma_z{\cal K}$, and ${\cal C}=\sigma_z$, with ${\cal K}$ complex conjugation, such that
\begin{equation}\label{eq:ph_tr_chiral}
 \begin{split}
  {\cal H}({\bf k}) =& {\cal H}^*(-{\bf k}), \\
  \sigma_z{\cal H}({\bf k})\sigma_z =& -{\cal H}^*(-{\bf k}), \\
  \sigma_z{\cal H}({\bf k})\sigma_z =& -{\cal H}({\bf k}).
 \end{split}
\end{equation}

We choose a driving protocol which involves the cyclic modulations of the hopping amplitudes $J_i$.\cite{Kitagawa2010} Each of the three hoppings contains a uniform term $J_u$ which is independent of time $\tau$, as well as a term $J_s$ which is periodically varied in a stroboscopic manner. The three-step driving protocol reads:
\begin{enumerate}\label{eq:driving_protocol}
\item $J_x = J_s + J_u$, $J_{y,z}=J_u$ 

for $nT<\tau\leq nT+ T/3$,\\
\item $J_y = J_s + J_u$, $J_{z,x}=J_u$ 

for $nT+T/3<\tau\leq nT+ 2T/3$,\\
\item $J_z = J_s + J_u$, $J_{x,y}=J_u$ 

for $nT+2T/3<\tau\leq nT+ T$,
\end{enumerate}
with $T$ the driving period.

In the following we set $T=3$ and $\hbar=1$ expressing the Floquet operator as the product
\begin{equation}\label{eq:floquet_phs}
 {\cal F} = \exp( -i {\cal H}_3) \exp( -i {\cal H}_2) \exp( -i {\cal H}_1),
\end{equation}
where ${\cal H}_i$ are the Hamiltonians during the three steps of the driving protocol shown above. Even though at every instance of time the Hamiltonian \eqref{eq:ph_ham} is time-reversal symmetric, the sequence in which the $x$, $y$, and $z$-type hoppings are modulated implies that the Floquet operator \eqref{eq:floquet_phs} has broken time-reversal symmetry.

\begin{figure}[tb]
\includegraphics[width=0.95\columnwidth]{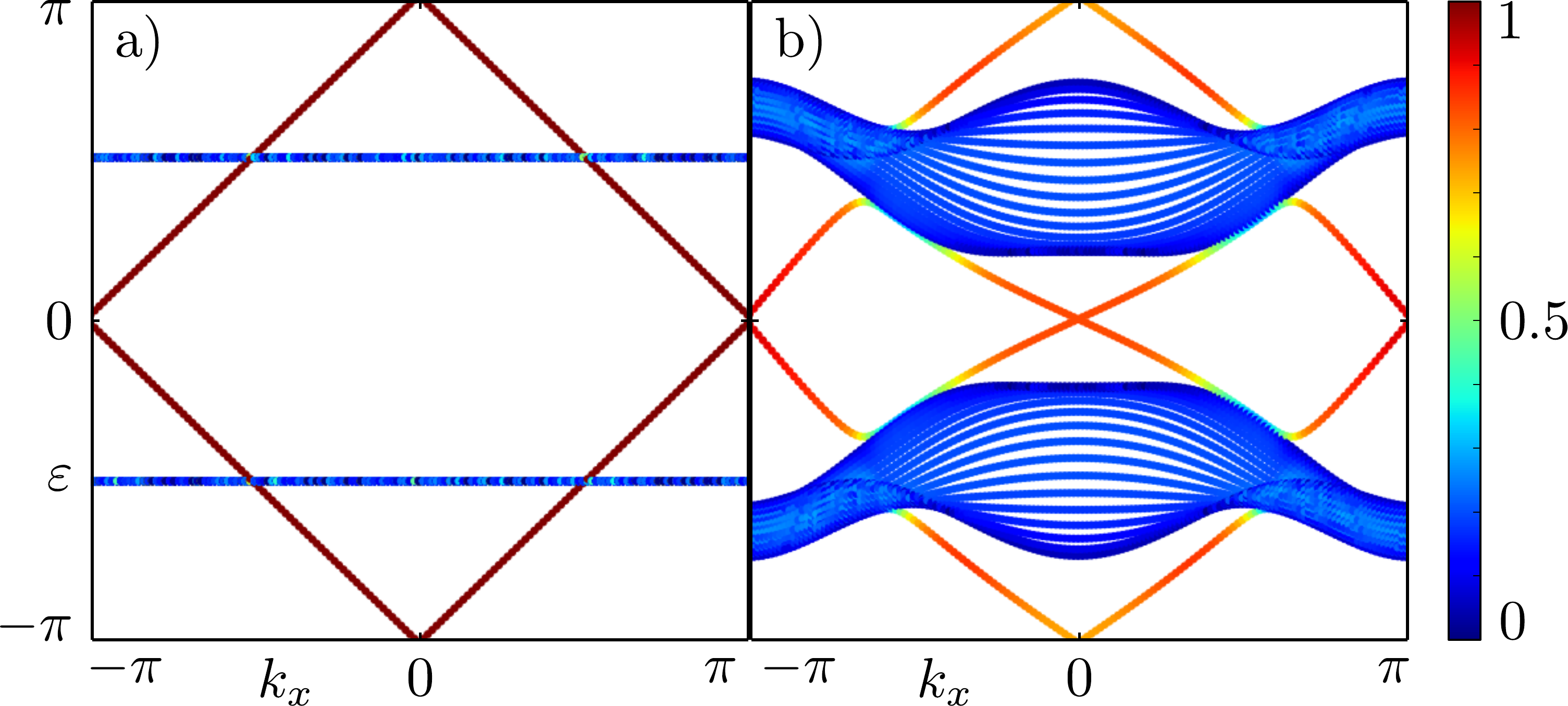}
\caption{Bandstructures of the model \eqref{eq:ph_ham} at $J_s=\pi/2$ in a strip geometry (infinite along $a_x$, $W=20$). In panel (a), $J_u=0$ and there are two flat bulk bands at quasi-energies $\varepsilon=\pm\pi/2$ as well as one chiral edge mode present at all quasi-energies. In panel (b) we set $J_u=0.25$, such that there is one chiral mode at $\varepsilon=\pi$ and a pair of counter-propagating modes at $\varepsilon=0$. The strong and weak invariants at $\varepsilon=\pi$ read $(C_{\rm{SM}}, \nu_{x,\pi}, \nu_{y,\pi})=(-1,1,1)$ in both panels. At $\varepsilon=0$ the system in panel (a) is in a $(-1, -1, -1)$ phase, while panel (b) has $(0, -1, -1)$. The color scale denotes the eigenstate intensity on the first and last $10\%$ of lattice sites. \label{fig:phs_bands}}
\end{figure}

In the simplest case we set $J_u=0$, such that during each of the three steps of the driving protocol only one of the $J_{x,y,z}$ hoppings is non-zero. When additionally $J_s=\pi/2$ a particle is transferred with unit probability between neighboring sites. Therefore, particles in the bulk fully encircle one hexagonal plaquette in two driving periods, leading to the formation of dispersionless (flat) bulk bands. In the presence of a boundary, the same driving protocol leads to the formation of a chiral propagating mode on the edge of the lattice, as shown in Fig.~\ref{fig:driving_protocol}. We recover the dispersionless bulk bands as well as the chiral edge modes in the quasi-energy banstructure of the system (see Fig.~\ref{fig:phs_bands}a). 

The emergence of the flat bulk bands attracts significant attention on its own right, in the context of possible realizations of exotic many-body phases in such systems.\cite{bergholtz_review2013, Parameswaran2013816,maksymenko_review2015, Parameswaran2013816, Maksymenko:2015aa} In the case of a non-trivial topology of the flat band, reflected in non-zero band Chern number, chiral edge states can emerge at the boundary of a finite system as a result of bulk-boundary correspondence. 

Interestingly, here for $J_u=0$ and $J_s=\pi/2$ the bulk Floquet operator, ${\cal F}=i(\cos(k_x) \sigma_x + \sin(k_x)\sigma_y)$ is independent of $k_y$, meaning that both of the flat bands have zero Chern number and all bulk states are trivially localized. The topological protection of the edge states can nevertheless be deduced from the topological invariant formulated in Ref.~\onlinecite{Rudner2013}, which takes into account the full time evolution of the Hamiltonian \eqref{eq:ph_ham} throughout the driving cycle.

In the following we show that scattering theory can not only reproduce this invariant, but reveals a much richer topological structure. To this end, we consider finite systems of size $L\times W$ (see Fig.~\ref{fig:driving_protocol}), where $L$ denotes the number of vertical bonds in the $a_x$ direction and $W$ the number of zig-zag chains in the $a_y$ direction. The scattering matrix associated to the Floquet operator is obtained by introducing absorbing terminals on the first and last zig-zag chain.

We repeat the analysis of Section \ref{sec:chiral}, finding that the presence of a chiral edge mode leads to a quantized transmission $G=1$. As before, we apply twisted boundary conditions (twist angle $\phi$) in the $a_x$ direction and express the strong topological invariant as the winding number of $\det\, r$, Eq.~\eqref{eq:rchern}. Remarkably, the scattering matrix invariant originally developed to compute the Chern number of static systems correctly describes this phase, even though all bulk bands are trivial. We find that the scattering matrix Chern number equals $C_{\rm{SM}}=-1$ for all quasi-energies $\varepsilon$ in the bulk gaps of Fig. \ref{fig:phs_bands}a.

Beyond the topological classification in terms of the strong invariant, the Hamiltonian \eqref{eq:ph_ham} shows a richer structure due to the presence of particle-hole symmetry. The cyclic nature of the driving protocol leads to a Floquet operator \eqref{eq:floquet_phs} which breaks time-reversal symmetry, but shows particle-hole symmetry (PHS) of the form

\begin{equation}\label{eq:ph_sym_floquet}
 {\cal F}({\bf k})=\sigma_z {\cal F}^*(-{\bf k}) \sigma_z.
\end{equation}

As such, for every eigenstate at quasi-energy $\varepsilon$ and momentum ${\bf k}$ there must be a state at $-\varepsilon$ and $-{\bf k}$. A chiral edge mode present at $\varepsilon=0,\pm\pi$ must therefore exist at points where ${\bf k}=-{\bf k}$. This is the case for the bandstructure shown in Fig.~\ref{fig:phs_bands}a, where the edge state crosses $k_x=0$ for $\varepsilon=\pi$ and $k_x=\pi$ for $\varepsilon=0$. The two edge mode configurations at $\varepsilon=0$ and $\varepsilon=\pi$ cannot be continuously deformed into each other without breaking particle-hole symmetry or closing the bulk gap, signaling the two phases are topologically distinct.

In other words, the parity of edge modes at momenta $k_x=0,\pi$ appearing in Fig.~\ref{fig:phs_bands}a is topologically protected. In a static topological superconductor this protection would be expressed in terms of a weak invariant.\cite{Asahi2012, Seroussi2014, Buehler2014, Diez2014a} Motivated by this fact, we analyze the gapped phases of the model \eqref{eq:floquet_phs} in terms of the scattering matrix invariants developed to classify weak topological superconductors.

To identify how the invariant emerges we analyze the symmetries constraining the scattering matrix. Plugging the particle-hole symmetry constraint \eqref{eq:ph_sym_floquet} into the scattering matrix expression \eqref{eq:FloquetS} leads to a relation
\begin{equation}\label{eq:ph_sym_S}
 S(\varepsilon) = \sigma_z S^*(-\varepsilon)\sigma_z,
\end{equation}
where $\sigma_z$ acts on the sub-lattice degree of freedom in the absorbing terminals.
As a consequence of Eq. \ref{eq:ph_sym_S}, at the particle-hole symmetric quasi-energies $\varepsilon=0,\pi$ there exists a basis in which the scattering matrix, and therefore also its reflection sub-block are real. We choose the basis $\tilde{r} = U r U^\dag$, with $U = {\rm diag}(1,i,1,i,\ldots,1,i)$ such that $\tilde{r}=\tilde{r}^*$, and introduce $\mathbb{Z}_2$ weak scattering matrix invariants counting the parity of edge modes at $k_x=0$ and $\pi$,

\begin{equation}\label{eq:phs_weak_inv}
\begin{split}
\nu_{x,0} &= {\rm sign}\,\det \tilde{r}(\phi=0), \\
\nu_{x,\pi} &= {\rm sign}\,\det \tilde{r}(\phi=\pi).
\end{split}
\end{equation}

Here, $\phi=0,\pi$ corresponds to applying periodic or anti-periodic boundary conditions in the $a_x$ direction, respectively. The weak invariants in the $a_y$ direction, $\nu_{y,0/\pi}$, can be defined in a similar manner, by introducing absorbing terminals at $a_x=1,L$ and using (anti-)periodic boundary conditions along $a_y$.\cite{oddsites}

The topological numbers of Eq.~\eqref{eq:rchern} and \eqref{eq:phs_weak_inv} have features similar to those defined for static systems. The strong $\mathbb{Z}$ index counts the net number of chiral modes, whereas the weak $\mathbb{Z}_2$ invariants determine the parity of modes crossing at momentum $0$ or $\pi$. As such, the weak invariants are constrained by the parity of the strong index $C_{\rm{SM}}$ as $\nu_{i,0}\nu_{i,\pi}=(-1)^{C_{\rm{SM}}}$, with $i=x,y$.\cite{R2010} Characterizing a topological phase therefore requires specifying $C_{\rm{SM}}$, $\nu_{x,\pi}$, as well as $\nu_{y,\pi}$, since their values can change independently.\cite{Diez2015} In accord with the terminology used in static weak topological superconductors, we label a phase where an odd number of edge modes cross momentum $\pi$ as non-trivial.\cite{Seroussi2014, setplusone}

For $J_u=0$ and $J_s=\pi/2$ we find that in both directions $i=x,y$ we have $\nu_{i,0}=-1$ and $\nu_{i,\pi}=1$ at $\varepsilon=\pi$, while their signs are reversed at $\varepsilon=0$. Therefore, the $\varepsilon=0$ phase of Fig.~\ref{fig:phs_bands}a is non-trivial in both a strong and a weak sense. Even though the strong invariant $C_{\rm{SM}}=-1$ remains unchanged, weak indices can distinguish between the two phases, such that an interface between the phase at $\varepsilon=0$ and $\varepsilon=\pi$ would host a pair of counter-propagating edge modes.\cite{Diez2015}
A purely weak phase is obtained in Fig.~\ref{fig:phs_bands}b, where turning on a small uniform hopping, $J_u=0.25$, changes the strong invariant to $C_{\rm{SM}}=0$, while keeping $\nu_{i,\pi}=-1$ at $\varepsilon=0$.

As in time-independent topological superconductors, the weak invariants introduced above are protected by the combination of PHS and translational symmetry of the system. Breaking the latter, for instance by introducing a staggered modulation of hopping amplitudes, would fold the quasi-BZ of Fig.~\ref{fig:phs_bands} in momentum, converting the non-trivial weak invariants at $\varepsilon=0$ into trivial ones.
Nevertheless, we find that weak invariants are robust to disorder as long as the latter preserves translational symmetry on average, and the $\varepsilon=0$ phase in Fig.~\ref{fig:phs_bands}b can be thought of as the periodically driven analogue of a statistical topological insulator once disorder is introduced.\cite{Fulga2014} In this phase, we verify that adding a random component to $J_u$, drawn independently for each bond from a uniform distribution, leads to an algebraically decaying edge transmission $G\sim1/\sqrt{L}$. The counter-propagating edge states avoid localization due to a combination of particle-hole symmetry and average translational symmetry, effectively forming a Floquet Kitaev edge.\cite{Diez2014a}

There are however important differences between the topological classification of time-independent superconductors and that of the model \eqref{eq:floquet_phs}. First, in periodically driven systems there are two PHS quasi-energies at which the weak invariants can be defined, $\varepsilon=0$ and $\pi$, allowing for a richer topological structure. Even though the weak invariants at $\varepsilon=0,\pi$ are protected by the same set of symmetries, PHS and translation along either $a_x$ or $a_y$, they can vary independently of each other. This is to be contrasted to the topological classification of static systems,\cite{Kitaev2009, Schnyder2009, Diez2015} in which different invariants always require different sets of protecting symmetries.

\begin{figure}[tb]
 \includegraphics[width=0.95\columnwidth]{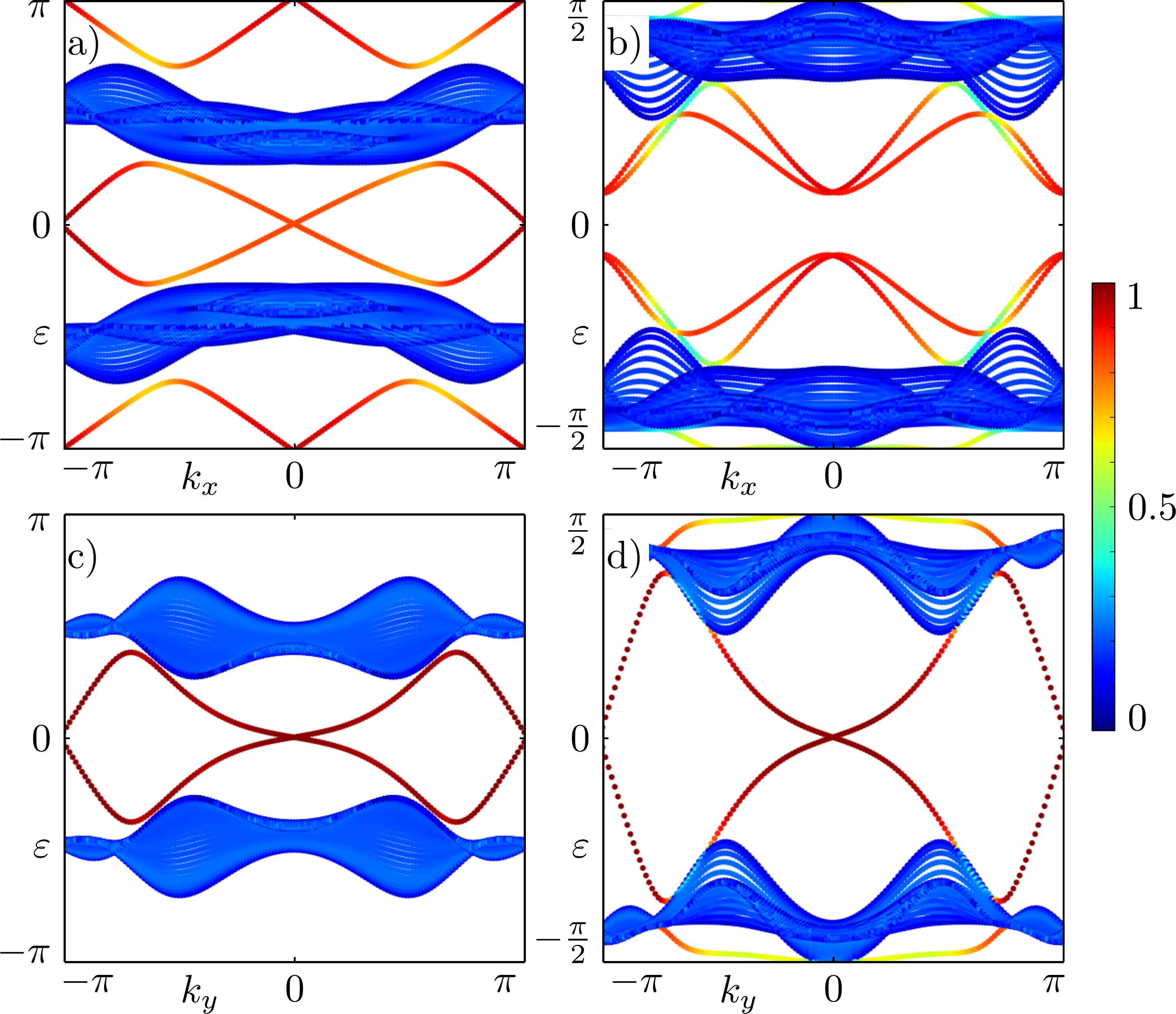}
 \caption{Bandstructure of the model \eqref{eq:ph_ham} in an infinite strip geometry with boundaries along $a_x$ (top, $W=30$) and $a_y$ (bottom, $L=30$) directions. We use $J_s=\pi/2$ and inequivalent uniform components of the hopping amplitudes, $J_{u,x}=-0.05$, $J_{u,y}=J_{u,z}=0.5$. In the right panels the BZ is folded in quasi-energy by changing the stroboscopic part of $J_x$ to $J_{s,x}=\pi/2+0.6$ on every second period. The invariants of the pre-folding phases (left panels), $(C_{\rm{SM}}, \nu_{x,\pi}, \nu_{y,\pi})=(0, -1, -1)$ at $\varepsilon=0$ and $(0, -1, 1)$ at $\varepsilon=\pi$ combine to form a new gapped phase with $(0, 1, -1)$ at $\varepsilon=0$. The color scale denotes the eigenstate intensity on the first and last $10\%$ of lattice sites. \label{fig:phs_weakbands}}
\end{figure}

Another feature unique to weak topological Floquet systems can be traced back to the periodicity of quasi-energy. In static systems a weak phase may be destroyed by a dimerization induced breaking of translational symmetry, leading to a doubling of the unit cell and a folding of the BZ in momentum. In periodically driven systems there is an extra direction in which the BZ can be folded, that of quasi-energy. Such a breaking of \emph{time translation} symmetry can be achieved by introducing a half-frequency component in the driving protocol, effectively doubling the driving period. This leads to a new phase at $\varepsilon=0$, with invariants $(C_{\rm{SM}}, \nu_{x,\pi}, \nu_{y,\pi})$ obtained by the composition of indices in the original (pre-folding) $\varepsilon=0$ and $\varepsilon=\pi$ phases. While in our example the folded phase is obtained simply by the $\mathbb{Z}$ addition of the strong index and the $\mathbb{Z}_2$ addition of the weak ones, more complex composition rules may apply in a generic setting.\cite{R2010}

We show an example of time folding in Fig.~\ref{fig:phs_weakbands}, obtained using inequivalent hopping amplitudes $J_{x,y,z}$ in Eq.~\eqref{eq:ph_ham}. Setting $J_s=\pi/2$ and the uniform components to $J_{u,x}=-0.05$ and $J_{u,y}=J_{u,z}=0.5$, we obtain phases with indices $(0, -1, -1)$ at $\varepsilon=0$ and $(0, -1, 1)$ at $\varepsilon=\pi$. The folding is implemented by repeating the three-step driving protocol described above twice, and altering the stroboscopic component of $J_x$ during every second period, as $J_{s,x}=J_s + \delta J_s$, with $\delta J_s=0.6$. This leads to a new gapped phase at $\varepsilon=0$, with invariants $(0, 1, -1)$, obtained by the composition of the original ones.

\section{Conclusion}
\label{sec:conc}

In many static systems, topological properties are fixed during the fabrication process. Part of the active interest in Floquet systems comes from the possibility to tune these properties by altering the driving protocol. Nevertheless, the topological classification of periodically-driven systems has remained largely unexplored.

We have shown that topological phases of periodically-driven systems can be analyzed in a unified framework, by using scattering theory. In some cases, both static and Floquet systems may be characterized by the same scattering matrix invariants, owing to the similar transport properties of their edge modes. This has enabled us to reveal a richer topological structure in a previously studied model (Section \ref{sec:phs}). We wrote down expressions for both strong and weak topological invariants, showing they correctly describe the non-trivial phases even when all bulk bands are trivial. 

When driven systems show weak phases for which no time-independent counterpart is known, as in Section \ref{sec:chiral}, scattering theory can be readily used to formulate novel invariants, based on the constraints imposed on the scattering matrix by the symmetries of the system. Since scattering matrix invariants are naturally tailored to the study of disordered systems, we were able to use their expressions to deduce the conditions under which the weak phase survives disorder.

Finally, we have shown that breaking translational symmetry in time can be used to alter the topological invariants of a phase. This novel feature is unique to Floquet topological insulators and allows a greater level of control in the design and manipulation of topological phases. Using the model of Section \ref{sec:phs} we have demonstrated this idea and altered the topological invariants of a weak phase by doubling the driving period.

We emphasize that, even though we have focused on two specific models, our scattering matrix approach to Floquet systems is completely general. It can, for instance, also be applied to time reversal invariant Floquet topological insulators or higher dimensional systems, as was the case for the scattering matrix invariants of static systems.\cite{Fulga2012} Our method is numerical in nature and requires systems large enough to avoid finite-size effects and show a well-defined bulk mobility gap. However, given a system size, Floquet scattering matrix invariants are more efficient to compute than ones based on wavefunctions. No knowledge of Floquet eigenstates or time-integration is required, and the biggest computational cost comes from inverting the time-evolution operator in Eq.~\eqref{eq:FloquetS}.

We hope our work will pave the way towards a full classification of topological phases in periodically-driven systems, which would parallel that of time-independent Hamiltonians. Additionally, our work will prove useful in determining the robustness to disorder of non-trivial phases, an essential ingredient for their experimental realization.

\acknowledgments

The authors thank B. Tarasinski, J. M. Edge, A. G. Grushin and A. R. Akhmerov for helpful discussions. We acknowledge E. Berg, A. Stern and M.-T. Rieder for useful comments and collaboration on closely related projects. This work was supported by the European Research Council under the European Union's Seventh Framework Programme (FP7/2007-2013) / ERC Project MUNATOP, ERC synergy UQUAM project, the US-Israel Binational Science Foundation, ISF grant 1291/12 and the Minerva Foundation. 

\bibliography{flwti}

\begin{thebibliography}{78}
\bibitem [{odd()}]{oddsites}%
  \BibitemOpen
  \href@noop {} {}\bibinfo {note} {As in static systems, computing the weak
  invariant requires an $L\times W$ lattice which extends over an odd number of
  both $A$ and $B$ sites. This avoids opening a mini-gap in the edge state
  spectrum. We use an odd $L$ to compute $\nu_x$, and an odd $W$ for
  $\nu_y$.}\BibitemShut {Stop}%
\bibitem [{set()}]{setplusone}%
  \BibitemOpen
  \href@noop {} {}\bibinfo {note} {Note that though the sign of ${\rm det}\,r$
  is topologically protected, choosing the value $+1$ to represent a trivial
  system requires a specific basis choice for the scattering matrix, as in
  static systems.}\BibitemShut {Stop}%
\end{thebibliography}

\end{document}